\documentclass{article}
\usepackage{spconf,amsmath,graphicx}
\usepackage{hyperref}
\usepackage{array}
\usepackage{color}
\usepackage{booktabs}
\usepackage{dcolumn}
\usepackage{url}
\usepackage{siunitx}
\usepackage{subcaption} 
\usepackage{ragged2e}
\usepackage{cite}
\usepackage{multirow, multicol}
\usepackage{makecell}

\newcolumntype{C}[1]{>{\centering\arraybackslash}m{#1}}

\newcolumntype{M}[1]{>{\centering\arraybackslash}m{#1}}
\newcolumntype{K}[1]{D{/}{/}{#1}} 

\title{Dynamic-SUPERB: Towards A Dynamic, Collaborative, and Comprehensive Instruction-Tuning Benchmark for Speech}
\name{\begin{tabular}[c]{@{}c@{}c@{}c@{}c@{}c@{}}
      Chien-yu Huang$^{1}$, Ke-Han Lu$^{*1}$, Shih-Heng Wang$^{*1}$, Chi-Yuan Hsiao$^{\dagger 1}$, Chun-Yi Kuan$^{\dagger 1}$, Haibin Wu$^{\dagger 1}$ \\ 
      Siddhant Arora$^{\mathsection 2}$, Kai-Wei Chang$^{\mathsection 1}$, Jiatong Shi$^{2}$, Yifan Peng$^{2}$, Roshan Sharma$^{2}$, Shinji Watanabe$^{2}$ \\ Bhiksha Ramakrishnan$^{2,3}$, Shady Shehata$^{3}$, Hung-yi Lee$^{1}$
      \end{tabular}
      \thanks{$^*$co-second authors. \ $^{\dagger}$co-third authors. \ $^{\mathsection}$co-fourth authors.}
      }
\address{\begin{tabular}[c]{@{}c@{}}
     $^{1}$National Taiwan University, Taiwan, 
     $^{2}$Carnegie Mellon University, USA \\
     $^{3}$Mohamed bin Zayed University of Artificial Intelligence, United Arab Emirates
\end{tabular}}

\begin{document}
\ninept
\maketitle
\begin{abstract}
Text language models have shown remarkable \textit{zero-shot} capability in generalizing to unseen tasks when provided with well-formulated instructions.
However, existing studies in speech processing primarily focus on limited or specific tasks.
Moreover, the lack of standardized benchmarks hinders a fair comparison across different approaches.
Thus, we present Dynamic-SUPERB, a benchmark designed for building universal speech models capable of leveraging \textit{instruction tuning} to perform multiple tasks in a zero-shot fashion.
To achieve comprehensive coverage of diverse speech tasks and harness instruction tuning, we invite the community to collaborate and contribute, facilitating the \textit{dynamic} growth of the benchmark.
To initiate, Dynamic-SUPERB features \textit{55 evaluation instances} by combining 33 tasks and 22 datasets.
This spans a broad spectrum of dimensions, providing a comprehensive platform for evaluation.
Additionally, we propose several approaches to establish benchmark baselines.
These include the utilization of speech models, text language models, and the multimodal encoder.
Evaluation results indicate that while these baselines perform reasonably on seen tasks, they struggle with unseen ones.
We release all materials to the public and welcome researchers to collaborate on the project, advancing technologies in the field together\footnote{\label{fn:official-page}{\scriptsize\url{https://github.com/dynamic-superb/dynamic-superb}}}.
\end{abstract}
\section{Introduction}
\label{sec:intro}

Self-supervised learning (SSL) significantly improves the scalability of machine learning models and their ability to generalize across a wide range of tasks \cite{oord2018representation, baevski2020wav2vec, hsu2021hubert, mohamed2022self}.
When applying an SSL model to a downstream task, a common approach is to build a task-dependent network on top of the SSL model and fine-tune it with task-specific data.
However, this approach demands that users construct a unique model for each task.
As downstream tasks increase, this process becomes more resource-intensive and time-consuming.
To address these issues, several parameter-efficient tuning approaches have thus gained popularity, such as prompting \cite{chang2022speechprompt, chang2023speechprompt, wu2023speechgen, li2021prefix, liu2022p, liu2023gpt, lester2021power, shin2020autoprompt, peng2023prompting} and adapters \cite{chen2023exploring, gong2023listen, fu2022adapterbias, he2021effectiveness}.
Particularly, in natural language processing (NLP), \textit{instruction tuning} involves fine-tuning language models (LM) on datasets where the tasks are defined by specific instructions \cite{wei2021finetuned}.
This approach considerably amplifies the \textit{zero-shot} learning capacity of LMs, enabling them to efficiently handle unseen tasks.
However, it remains largely unexplored in the speech-processing field, where the aspects are even more abundant and diverse.
While there are existing studies like SPECTRON \cite{nachmani2023lms} and AudioPaLM \cite{rubenstein2023audiopalm} that focus on jointly processing text and speech, they are designed for specific tasks.
On the other hand, AudioGPT \cite{huang2023audiogpt}, though driven by text instructions, is limited to a pre-defined set of tasks.

This paper presents Dynamic-SUPERB, the first collaborative benchmark for instruction-tuning speech models.
Although previous benchmarks provide evaluations on several aspects, they are \textit{static} \cite{nguyen2020zero,yang21c_interspeech,shon2022slue,evain21_interspeech,gandhi2022esb,conneau2023fleurs,shi2023ml,lin2023utility}.
In contrast, Dynamic-SUPERB encourages the community to contribute a broader range of tasks so that the task variations are \textit{dynamically} extended, aiming for a more comprehensive assessment.
For the first version, we gathered over 20 publicly available English datasets and transformed them into a diverse set of tasks.
These tasks span 6 dimensions: content, speaker, semantics, degradation, paralinguistics, and audio (non-speech).
In Dynamic-SUPERB, each task is basically composed of three parts: (I) text instructions, (II) speech utterances, and (III) text labels.
Following this, the system receives both the text instruction and speech utterances as input, and then functions based on the instruction.
For example, given the instruction ``Identify the emotion conveyed in the utterance'', the model performs emotion recognition and outputs the label ``Happy'' textually.
For simplicity in evaluation, we currently focus on classification tasks, such as intent classification and speaker verification, and leave generative ones for future collaboration with the community.
To enhance user engagement, we offer detailed documentation and establish a clear process for submitting new tasks to Dynamic-SUPERB.
All submitted tasks are subject to a review before inclusion.
Reviews mainly focus on technical accuracy and completeness, ensure clarity of the proposal, and determine if the task aptly evaluates its intended objectives.

We propose five approaches to establish baselines in Dynamic-SUPERB.
We integrated text embeddings from BERT \cite{devlin-etal-2019-bert} into the generative spoken language model (GSLM) \cite{lakhotia2021generative}, enabling operations on both speech and text.
We also adapted Whisper \cite{radford2023robust}, which was primarily designed for speech recognition involving both speech and text, to instruction tuning.
Besides modifying speech models, we integrated speech representations from either Whisper or ImageBind \cite{girdhar2023imagebind} into LLaMA \cite{touvron2023llama, touvron2023llamatwo}, a prevalent large language model (LLM).
We also utilize Whisper (ASR) and ChatGPT \cite{openai2023chatgpt} to build a concatenative system (ASR-ChatGPT).
Due to the lack of instruction-based tasks in the field, we developed Dynamic-SUPERB-Train using academic resources as a preliminary training set for the baselines.
Its overlap with Dynamic-SUPERB differentiates tasks into \textbf{seen} and \textbf{unseen} categories, assessing model generalizability.
Evaluation results indicate that no single model dominates across all tasks.
ASR-ChatGPT excels in some semantic tasks but underperforms with speaker and paralinguistic tasks.
Speech models that integrate text struggle with unseen tasks due to their limited comprehension of text instructions.
Conversely, combining speech representations with a text language model surpassed other methods.

\section{Dynamic-SUPERB}
\label{sec:dynamic-superb}

\subsection{Objectives}
\label{sec:objective}
Dynamic-SUPERB is committed to offering a comprehensive evaluation of universal speech models and facilitating advancements in this field.
We believe that instruction tuning is a pivotal step towards universal speech models.
While we included a wide range of tasks to kickstart the benchmark, there are always numerous innovative and creative applications in speech processing.
Unlike previous benchmarks with a fixed task set \cite{nguyen2020zero,yang21c_interspeech,shon2022slue,evain21_interspeech,gandhi2022esb,conneau2023fleurs,shi2023ml,lin2023utility}, Dynamic-SUPERB seeks to dynamically expand its task variety through community collaboration, just as its name indicates.
We offer a straightforward pipeline for easy user engagement and have established a review guideline outlining how the community can contribute to Dynamic-SUPERB.

\subsection{Terminologies}
\label{sec:term}
We collected several datasets and used them in a variety of tasks.
For example, we can utilize LibriSpeech \cite{panayotov2015librispeech} for speaker verification (multi-speaker dataset), and spoken term detection (transcription available), to name just a few applications.
Therefore, in Dynamic-SUPERB, a dataset can correspond to multiple tasks, and a task may cover several datasets.
In this paper, a \textbf{task} is the specific type of processing or operation to be carried out, such as ``speaker verification'' or ``intent classification''.
It is a general category that defines the kind of work or operation being done, without specifying the dataset on which it is performed.
Meanwhile, an \textbf{instance} refers to the specific combination of a task and a dataset.
For example, performing ``speaker verification'' on the ``LibriSpeech'' dataset would be an instance.
This term describes the exact pairing of the task and dataset we evaluate or discuss.

\subsection{Tasks \& datasets}
\label{sec:tasks}
Generally, tasks can be categorized as either discriminative or generative.
Discriminative tasks aim to sort data into predefined class sets.
In contrast, generative tasks involve producing sequences of tokens or signals of variable lengths, making them inherently more challenging.
As an initial launch to simplify the evaluation process, we only consider classification tasks in the current version of Dynamic-SUPERB.
These tasks assess the capabilities of models across various dimensions, including content (\textbf{CNT}), speaker (\textbf{SPK}), semantics (\textbf{SEM}), degradation (\textbf{DEG}), and paralinguistics (\textbf{PRL}).
Besides, we added audio-processing tasks (\textbf{AUD}) to investigate how baseline models extend beyond speech processing.
Table~\ref{tab:task-summary} outlines the tasks in Dynamic-SUPERB by the above-mentioned 6 dimensions.

\begin{table}[ht]
    \centering
    \caption{
    Statistics of Dynamic-SUPERB.
    For a detailed explanation of the abbreviations used, please refer to Sec.~\ref{sec:tasks}.
    }
    \label{tab:task-summary}
    \setlength{\tabcolsep}{1.5pt}
    \small
    \resizebox{.92\linewidth}{!}{
    \begin{tabular}{C{6em} C{3em}
        C{2.5em}C{2.5em}C{2.5em}
        C{2.5em}C{2.5em}C{2.5em}
    }
        \toprule
        \textbf{Attribute} & \textbf{Status} & \textbf{CNT} & \textbf{SPK} & \textbf{SEM} & \textbf{DEG} & \textbf{PRL} & \textbf{AUD} \\
        \midrule
        Datasets & - & 5 & 3 & 3 & 6 & 7 & 3 \\
        \midrule
        \multirow{2}{*}{Tasks} & Seen & 3 & 2 & 2 & 4 & 3 & 0 \\
                & Unseen &2& 1 & 2 & 8 & 3 & 3 \\
        \midrule
        \multirow{2}{*}{Instances} & Seen & 9 & 3 & 2 & 6 & 4 & 0 \\
        & Unseen & 2 & 2 &4 & 13 & 3 & 7 
        \\
        \bottomrule
    \end{tabular}
    }
\end{table}

Given the lack of instruction-tuning data in the field, we collected Dynamic-SUPERB-Train to train baselines.
It shares some tasks with Dynamic-SUPERB, termed \textbf{seen tasks}, but uses different datasets for instances.
Due to the space limit, we cannot list all tasks in detail.
The full collection of tasks is available on the Dynamic-SUPERB official page\footref{fn:official-page}.

\subsection{Task format}
Each task consists of at least three components: speech utterance, instruction, and label.
We sourced it from widely-used corpora such as LibriSpeech \cite{panayotov2015librispeech}, LJSpeech \cite{ljspeech17}, and VCTK \cite{yamagishi2019vctk}.
Depending on the task specification, there may be more than one utterance.
For example, in speaker verification, two utterances are involved to determine whether they are produced by the same speaker.
For instructions, we chose to use text format rather than spoken format.
Spoken instructions, with varying content, speaker characteristics, and prosody, are more complex than text-based ones.
Text instructions serve as an intermediary, bridging text and spoken instruction tuning.
To generate various instructions, we initially created some basic ones manually and then used ChatGPT to rephrase them and generate additional variations.
This results in each task containing approximately 10 to 30 different types of instructions.

A standard classification model produces a distribution over all potential labels.
For instance, an emotion classifier assigns probabilities to each emotion category, including happy, angry, and sad.
The method is simple and efficient, but it constrains the adaptability of the model.
Adding a new emotion, like ``tired,'' would require retraining the model, let alone adapting it for entirely different tasks.
To tackle this issue, we believe that a universal speech model should produce outputs generatively.
\textit{Consequently, in Dynamic-SUPERB, all labels are represented textually.}
To explore generalizability on unseen tasks, all baseline models are designed to respond in a generative manner.
That is, without any constraints, a model can produce any arbitrary response.
For example, a model might respond with synonyms instead of the exact desired answer, complicating the evaluation process.
To address this, we specify the available options directly in the instruction.
Specifically, we append the phrase ``The answer could be A, B, or C'' to the instruction, where A, B, and C represent the candidate answers.
If ``A'' is the correct answer, then only the response ``A'' is considered correct.
Responses like ``The answer is A'' or synonyms of ``A'' are considered incorrect.

\subsection{Benchmark extendibility}
Dynamic-SUPERB is a benchmark designed with forward-thinking principles.
As detailed in Sec.~\ref{sec:objective}, it has the capability to dynamically expand its task coverage.
In most conventional benchmarks, adding a new task is challenging because it requires users to undergo re-training specific to the new task.
However, Dynamic-SUPERB leverages instruction tuning, enabling a single model to handle multiple tasks \textit{without explicit fine-tuning}.
This feature significantly simplifies the process when new tasks are integrated, highlighting the extendibility of the benchmark.

Contrasting many existing benchmarks that focus on static evaluations, Dynamic-SUPERB adopts a more dynamic and flexible design.
Recognizing the ever-changing landscape of speech-processing tasks, we believe that benchmarks should evolve alongside the innovations and advances in research.
Though Dynamic-SUPERB may not encompass every potential speech-processing task at present, its real value lies in its design that encourages growth and the incorporation of emerging tasks.
This inclusive framework not only welcomes but thrives on community collaboration, facilitating the introduction of innovative tasks and methods.
Such extendibility ensures that Dynamic-SUPERB remains relevant and effective in the fast-paced world of speech-processing research.

\section{Baseline Frameworks}
\label{sec:baseline-frameworks}
While instruction tuning is a promising method for achieving universal speech models, there is no model that has been specifically designed to do so.
Therefore, based on existing models, we propose five approaches to instruction tuning and serve them as the baseline models.
These models were directly trained with Dynamic-SUPERB-Train and tested on Dynamic-SUPERB.

\subsection{BERT-GSLM}
\label{sec:instr-gslm}
GSLM \cite{lakhotia2021generative} comprises three components: a discrete feature extractor (utilizing HuBERT and k-means clustering), a unit language model (uLM), and a speech synthesizer.
The first two components are used here.
To adapt the GSLM for instruction tuning, we aim to enhance the uLM to handle both input and output in textual form.
For the input, we employ BERT \cite{devlin-etal-2019-bert} to derive contextualized representations of text tokens.
These are then projected into the GSLM embedding space using a linear layer.
For the output, originally, the uLM produced a probability distribution exclusively over speech tokens.
We expanded the token set by incorporating text tokens, allowing the uLM to produce a distribution that includes both speech and text tokens.
During training, feature extractors (HuBERT \& BERT) remained fixed, while the uLM and the projection layer were trainable.

\subsection{Whisper}
\label{sec:whisper}
Whisper \cite{radford2023robust} is a large-scale ASR model, but its encoder captures a wealth of information beyond mere content \cite{gong23d_interspeech}.
The decoder transcribes based on speech features and previously generated texts, making it potentially suitable for our scenario.
We add the instruction to the text input of the decoder, enabling it to autoregressively generate the desired output conditioned on the instruction.

In the original Whisper implementation, several special tokens were used to signify specific functions, and we also adhered to this for instruction tuning.
Text instructions are appended after the \texttt{PREV} token, followed by the start of transcription token, a language tag token, and a timestamp disabling token.
Whisper then auto-regressively generates a response based on these tokens and the encoder-extracted speech features.
During training, we employed Whisper-medium, and both the encoder and decoder were updated.

\subsection{Multi-modal large language model}
LLMs have demonstrated exceptional proficiency in instruction tuning tasks within NLP, prompting our choice of LLaMA \cite{touvron2023llama, touvron2023llamatwo} here.
Specifically, we used ImageBind-LLM \cite{han2023imagebind}, a multi-modal variant of LLaMA.
ImageBind \cite{girdhar2023imagebind} is a multi-modal encoder that embeds several data types, such as images and audio, into a unified space.
Speech samples are converted into utterance-level features and subsequently integrated with LLaMA through the zero-initialized injection mechanism \cite{han2023imagebind, gao2023llamaadapterv2}.
We employed LLaMA-Adapter \cite{zhang2023llamaadapter, gao2023llamaadapterv2} for computational efficiency.
During training, ImageBind remained fixed, and only adapter modules in LLaMA were updated.

While ImageBind-LLM shows promise, we observed that its utterance-level encoding overlooks the temporal nuances of speech.
Therefore we introduced Whisper-LLM to address this limitation.
We substituted ImageBind with the Whisper encoder to better capture temporal information.
The training process was similar to that of ImageBind-LLM, with the exception of an added downsampling step to reduce the sequence length of speech features.

\subsection{ASR-ChatGPT}
\label{sec:asr-chatgpt}
The above models require re-training.
We also adopted a concatenative framework using pre-trained models, as demonstrated in a previous study \cite{he2023can}.
Specifically, we used Whisper to transcribe the speech and then merge it with the instruction to guide ChatGPT in responding.
However, we found responses from ChatGPT vary.
For example, an answer ``A'' might be elaborated as ``The answer is A.''
To ensure ChatGPT adheres to our requirements, we added a prompt: ``Please select one label from the provided options and respond with that label only.''
Empirically, this constrains ChatGPT effectively.

\section{Results}
\label{sec:results}
In this section, we present a preliminary study of the baseline models.
Specifically, we first discuss how the models perform on seen tasks (Sec.~\ref{sec:seen-tasks} \& Sec.~\ref{sec:seen-unseen-instrs}), and then examine their capability of generalization towards unseen tasks (Sec.~\ref{sec:unseen-tasks}).

\begin{table}[t]
    \centering
    \caption{Average accuracy (\%) of the proposed baselines on seen tasks in Dynamic-SUPERB across different dimensions. Entries marked with ``-'' indicate an empty task set.}
    \label{tab:seen-tasks}
    \setlength{\tabcolsep}{1.5pt}
    \resizebox{.82\linewidth}{!}{
    \begin{tabular}{
    C{2.4cm}
    C{0.8cm}C{0.8cm}C{0.8cm}C{0.8cm}C{0.8cm}C{0.8cm}
    }\toprule
         \textbf{Model} & \textbf{CNT} & \textbf{SPK} & \textbf{SEM} & \textbf{DEG} & \textbf{PRL} & \textbf{AUD} \\
         \midrule
         BERT-GSLM & 66.3 & 49.1 & 47.2 & 68.2 & 52.7 & - \\
         Whisper & 95.3 & 47.9 & 55.5 & 71.1 & 49.4 & - \\
         ImageBind-LLM & 64.3 & 54.7 & 47.6 & 78.7 & 59.8 & - \\
         Whisper-LLM & 77.6 & 91.7 & 55.7 & 91.0 & 66.3 & - \\
         ASR-ChatGPT & - & - & - & - & - & - \\
         \midrule
         Random & 49.9 & 40.2 & 41.0 & 45.9 & 67.1 & - \\
         \bottomrule
    \end{tabular}
    }
\end{table}

\begin{table}[t]
    \centering
    \caption{Average accuracy (\%) of the proposed baselines on unseen tasks in Dynamic-SUPERB across different dimensions.}
    \label{tab:unseen-tasks}
    \setlength{\tabcolsep}{1.5pt}
    \resizebox{.82\linewidth}{!}{
    \begin{tabular}{
    C{2.4cm}
    C{0.8cm}C{0.8cm}C{0.8cm}C{0.8cm}C{0.8cm}C{0.8cm}
    }\toprule
         \textbf{Model} & \textbf{CNT} & \textbf{SPK} & \textbf{SEM} & \textbf{DEG} & \textbf{PRL} & \textbf{AUD} \\
         \midrule
         BERT-GSLM & 0.0 & 32.8 & 5.3 & 41.6 & 12.6 & 0.0 \\
         Whisper & 14.4 & 58.0 & 13.8 & 55.4 & 8.5 & 0.8 \\
         ImageBind-LLM & 15.7 & 45.4 & 24.7 & 47.6 & 20.6 & 35.7 \\
         Whisper-LLM & 8.7 & 60.6 & 20.9 & 59.0 & 6.6 & 15.9 \\
         ASR-ChatGPT & 65.0 & 40.1 & 69.3 & 43.5 & 22.9 & 9.8 \\
         \midrule
         Random & 11.8 & 50.2 & 33.1 & 43.1 & 21.0 & 23.4 \\
         \bottomrule
    \end{tabular}
    }
\end{table}

\subsection{Seen tasks}
\label{sec:seen-tasks}

Table~\ref{tab:seen-tasks} presents the evaluation results for each baseline on seen tasks in Dynamic-SUPERB.
Due to space constraints, we report the average accuracy for each dimension.
Detailed accuracy for each task is available on our website\footref{fn:official-page}.
We also included the random baseline for comparison, which was obtained by randomly selecting answers from the label distribution of each task.
Here, we discuss the performance on \textbf{seen tasks}, which were used in both training and testing.
ASR-ChatGPT is included for comparison (Table~\ref{tab:unseen-tasks}), even though the term ``seen tasks'' does not apply to it.
The audio (AUD) dimension is not discussed here because there are no seen tasks.

Overall, GSLM performed poorly.
While the best performance of GSLM was in the degradation dimension (DEG), its accuracy only surpassed that of ASR-ChatGPT.
We believe this is likely because GSLM has fewer parameters than other models, and using discrete speech tokens may discard crucial information required for these tasks.
Whisper showed an outstanding accuracy in the content dimension (CNT).
This is probably because Whisper is primarily an ASR model, contributing to its high accuracy in processing content information.
ImageBind-LLM had a slightly better performance than GSLM.
Despite using utterance-level embeddings, it achieved much higher accuracy than the random baseline in CNT, which we believed would require the use of temporal information.
This suggests that, as of now, some tasks in this dimension do not necessitate detailed temporal information.
We hope the community will contribute more diverse tasks to enrich it.
Whisper-LLM dominated in most dimensions except CNT.
Particularly, it achieved very high accuracy in the speaker (SPK) and degradation (DEG) dimensions.
The performance difference between Whisper and Whisper-LLM suggests that while the Whisper encoder provides abundant information, the way we utilize this information significantly impacts performance (comparing the Whisper decoder to LLaMA).
ASR-ChatGPT excelled in the semantics dimension (SEM) and had a much higher accuracy than all the other baselines.
Conversely, it failed in those involving several speech characteristics (SPK, DEG, and PRL) because the ASR process discarded the essential information.

\subsection{Seen \& unseen instructions in seen tasks}
\label{sec:seen-unseen-instrs}
In the seen tasks, certain instructions appeared in both training and testing (\textbf{seen instructions}), whereas others were exclusive to testing (\textbf{unseen instructions}).
We then analyzed how baseline models performed from the view of seen and unseen instructions.
Table~\ref{tab:instruction-summary} lists the performance comparison based on seen/unseen instructions.
AUD is excluded as it was not utilized during training.
All the baselines followed the same trend.
In CNT, SPK, and SEM, there was little difference between the seen and unseen instructions.
For DEG and PRL, the unseen instructions had a greater impact, leading to decreased accuracies.
In the two dimensions, there was a relatively serious imbalance between the seen and unseen instructions, so the average accuracy was influenced much by a specific task that models did not perform well on.
In our preliminary study, we endeavored to provide as many instructions as possible for each task.
However, in real-world scenarios, there can be hundreds or even thousands of ways to formulate instructions specific to a task.
Achieving such diversity is unlikely with just a small group of researchers or solely with tools like ChatGPT.
This underscores the importance of community collaboration to diversify and enhance the benchmark.

\begin{table}[t]
    \centering
    \caption{Average accuracy (\%) across different dimensions based on seen/unseen instructions. Only seen tasks were evaluated.}
    \label{tab:instruction-summary}
    
    \setlength{\tabcolsep}{1.7pt}
    \small
    \resizebox{.97\linewidth}{!}{
    \begin{tabular}{C{7em}C{5em}
        C{3em}C{3em}C{3em}
        C{3em}C{3em}C{3em}
    }
         \toprule
        \textbf{Model} & \textbf{Instruction} & \textbf{CNT} & \textbf{SPK} & \textbf{SEM} & \textbf{DEG} & \textbf{PRL} \\
         \midrule
         \multirow{2}{*}{BERT-GSLM} & Seen & 66.5 & 48.8 & 47.3 & 68.2 & 69.3 \\
                                    & Unseen & 65.9 & 49.2 & 46.9 &60.6 & 54.6 \\
        \midrule
         \multirow{2}{*}{Whisper} & Seen &  95.4 & 48.2 & 55.4 & 71.0 & 65.8 \\
                                 & Unseen & 95.0 & 47.3 & 55.7 & 50.5 & 49.6 \\
        \midrule
        \multirow{2}{*}{ImageBind-LLM} & Seen &  64.1 & 55.1 & 47.3 & 78.6 & 75.8\\
                                       & Unseen & 64.8 & 54.4 & 48.2 & 62.8 & 59.7\\
        \midrule
         \multirow{2}{*}{Whisper-LLM} & Seen & 77.3 & 91.7 & 55.3 & 91.1 & 84.7 \\
                                     & Unseen & 78.1  & 91.7 & 56.8 & 84.9 & 66.4 \\
         \bottomrule
    \end{tabular}
    }
\end{table}

\subsection{Unseen tasks}
\label{sec:unseen-tasks}
Table~\ref{tab:unseen-tasks} reveals a significant performance gap between seen and unseen tasks across all models.
While baseline models performed reasonably on seen tasks, their performance declined on unseen tasks.
In typical classification tasks, a random guess serves as the baseline for the lowest performance, but this is not the case here.
Although the instructions explicitly outlined available options, the model responded in a generative manner.
It could produce answers that were not among the given options, leading to a performance that was even worse than what a random guess would achieve.
BERT-GSLM and Whisper struggled to select from the provided options, which is evidenced by their accuracy falling below the expected random baseline.
Notably, BERT-GSLM had a very low accuracy of 0.0\% on unseen tasks in CNT.
On the other hand, ImageBind-LLM and Whisper-LLM seemed to make selections that appeared almost random from the instructions, which resulted in an accuracy that hovered around the random baseline.
These indicate that while LLMs might not always perform a specific task as instructed, they exhibit an understanding of how options are presented in natural language.
This proficiency likely stems from their pre-training on large-scale text data.
Conversely, speech models, without text pre-training and trained on limited data, could not achieve the same proficiency.

Additionally, we observed that most models underperformed in AUD.
This can be attributed both to its absence during the training and the intrinsic distinction between speech and audio signals.
Conversely, ImageBind-LLM exhibited superior accuracy because of its pre-training on several data modalities, including audio.

From our analysis of seen/unseen instructions (Sec.~\ref{sec:seen-unseen-instrs}) and tasks (this section), we surmise that these models could not genuinely perform specific tasks based on the semantic information in the instructions.
Instead, they often performed tasks by recognizing specific patterns in the instructions, such as a bag-of-words approach.
Within a task, despite variations in instructions, they might possess a similar bag of words due to their design aimed at specifying the same task.
On the other hand, different tasks have unique instructions (and patterns), and the model might fail if these were not used in training.
This explains why these models demonstrated notable performance on seen tasks with unseen instructions but faltered in generalizing effectively to unseen tasks.
The instructions for unseen tasks have very different patterns than those for seen tasks.

\section{Conclusions}
\label{sec:conclusions}
Instruction tuning is increasingly popular for enabling zero-shot applications in NLP, but it remains underexplored in speech processing.
This paper presents Dynamic-SUPERB, the first dynamic and collaborative benchmark for instruction tuning in speech models, offering a comprehensive exploration across diverse dimensions.
Five different approaches are proposed and tested on the benchmark, showing encouraging results on seen tasks.
However, their performance on unseen tasks highlights the need for continued research in this field.
We open-source all the materials to lower the barrier for reproduction, benchmarking, and analysis of instruction tuning in speech processing.
We welcome researchers to join our active community and drive the research frontier together.

\section{Acknowledgements}
We thank the National Center for High-performance Computing (NCHC) of the National Applied Research Laboratories (NARLabs) in Taiwan for providing computational and storage resources.

\bibliographystyle{IEEEbib}
\bibliography{strings,refs}

\end{document}